\documentclass[aps,prb,twocolumn,showpacs,superscriptaddress]{revtex4}

\usepackage{graphicx} 
\usepackage{soul}
\usepackage{multirow}
\usepackage[T1]{fontenc}
\usepackage[latin9]{inputenc}
\usepackage{rotating}
\usepackage{amsmath}
\usepackage{amssymb}
\usepackage{color}

\begin{document}

\title{Chlorine Adsorption on Graphene: Chlorographene }

\author{H. \c{S}ahin} \email [New address: Departement Fysica, Universiteit Antwerpen, Groenenborgerlaan 171, B-2020 Antwerpen, Belgium ] {}
\affiliation{UNAM-National Nanotechnology Research Center, Bilkent University, 06800 Ankara, Turkey}
\affiliation{Institute of Materials Science and Nanotechnology, Bilkent University, 06800 Ankara, Turkey}
\author{S. Ciraci} \email [ Corresponding author: ]{ciraci@fen.bilkent.edu.tr}
\affiliation{UNAM-National Nanotechnology Research Center, Bilkent University, 06800 Ankara, Turkey}
\affiliation{Institute of Materials Science and Nanotechnology, Bilkent University, 06800 Ankara, Turkey}
\affiliation{Department of Physics, Bilkent University, 06800 Ankara, Turkey}

\date{\today}

\begin{abstract}
We perform first-principles structure optimization, phonon frequency and finite 
temperature molecular dynamics calculations based on density functional theory 
to study the interaction of chlorine atoms with graphene predicting the 
existence of possible chlorinated graphene derivatives. The bonding of a single 
chlorine atom is ionic through the transfer of charge from graphene to
chlorine adatom and induces negligible local distortion in the underlying planar 
graphene. Different from hydrogen and fluorine adatoms, the migration of a single 
chlorine adatom on the surface of perfect graphene takes place almost without 
barrier. However, the decoration of one surface of graphene with Cl adatoms 
leading to various conformations cannot sustain due to strong Cl-Cl
interaction resulting in the desorption through the formation of Cl$_2$ 
molecules. On the contrary, the fully chlorinated graphene, chlorographene CCl, 
where single chlorine atoms are bonded alternatingly to each carbon atom from different 
sides of graphene with $sp^3$-type covalent bonds, is buckled. We found that this 
structure is stable and is a direct band gap semiconductor, whose band gap can be 
tuned by applied uniform strain. Calculated phonon dispersion relation
and four Raman-active modes of chlorographene are discussed.

\end{abstract}

\maketitle

\section{Introduction}

The synthesis of graphene\cite{novo1,novo2} has led to intense research activity
in the field of graphene-based nanoscale devices. Although graphene is one of the
most mechanically strong material having a wide range of extraordinary properties,
\cite{novo1,novo2,geim,berger,kats} practical device applications are limited by
its metallic behavior and sensitivity to surface adsorbates.

Efforts to synthesize chemically modified graphene composites with tailored electronic,
optical, and chemical properties have presented new directions in graphene
research. In particular, band gap engineering of graphene through chemical
modification, such as oxygenation,\cite{dikin, eda, navarro, gilje, robinson,kim} and
hydrogenation\cite{elias, flores, apl2009, prb2010, graphane-m, sofo} is
appealing for electronic applications, since the scalable fabrication of
graphene-based devices without disturbing the strong honeycomb lattice has
become possible. However, due to the complex atomic structure of
grapheneoxides\cite{eda} (GOs) and thermal instabilities of hydrogenated
graphenes (CHs) even at low temperatures,\cite{elias, flores, apl2009} search
for the novel graphene-based materials is still continuing.

Over the past three years experimental\cite{nair, cheng, robinson2, withers} and
theoretical\cite{leenaerts,hasancf} studies have demonstrated that
chemical conversion of graphene to fluorographene (CF) is possible. In addition to
early studies on the atomic composition and band structure of fluorocarbon
materials,\cite{charlier, takagi} it was reported that the monolayer CF has quite
different vibrational spectra and Raman characteristics as compared to hydrogenated
graphene analogues.\cite{leenaerts} We also investigated the electronic and
elastic properties of possible fluorinated graphene materials and attempted to
clarify the discrepancy between theoretical and experimental results.\cite{hasancf}
Easy synthesis, high-quality insulating behavior and extraordinary mechanical
strength of fluorographene (CF) have inspired intense research on other halogen
decorated graphene derivatives.

In addition to three known derivatives of graphene: GO, CH and CF, the successful
synthesis of chlorinated graphene was also achieved very recently.\cite{bli} It is
experimentally demonstrated that nondestructive and patternable conversion of
graphene is possible by using various photochemical chlorination
techniques.\cite{bli, new1, new2} While the research on chlorine-graphene
interaction is rapidly
growing,\cite{wehling,klintenberg,medeiros,zboril,rudenko,bli, new1,
new2, harvest,ijas,jpc} comprehensive research on the stability of various
chlorinated graphene structures and their resulting properties are sparse.

In this paper we present a detailed analysis of the interaction between chlorine
atom and graphene leading to the chlorination of graphene. Although the
possibility of covering graphene surfaces with chlorine atoms has been reported,
analyses of structural stability, electronic and magnetic properties as a
function of Cl coverage are lacking. Our main motivation is to reveal which
conformations of chlorinated graphene are stable and how these conformations
modify the properties of graphene. To this end we investigated the chlorination
of graphene starting from single Cl adsorption to full coverage leading to
chlorographene, namely CCl. At low coverage with diminishing Cl-Cl coupling, the
binding of Cl to graphene is significant, but adsorbed Cl atoms migrate on the
surface of graphene almost without an energy barrier.  We found that the nonbonding
chair conformation of chlorographene\cite{klintenberg,medeiros} (CCl) consisting
of a planar graphene sandwiched between widely spaced two planar Cl layers is unstable.
On the other hand, the covalently bonded chair conformation of the chlorographene (CCl)
is found to be stable at T=0 K and possibly at room temperature.
This latter conformation consists of buckled graphene sandwiched
between two planar Cl layers and is a nonmagnetic semiconductor
with 1.2 eV direct band gap. Our results reconcile the discrepancy between
the experimental study\cite{bli} obtaining semiconducting properties upon
the chlorination of graphene and theoretical studies predicting metallic
state.\cite{klintenberg, medeiros}

\section{Computational Methodology}\label{method}

To investigate mechanical, electronic magnetic properties of chlorinated
graphene we carried out first-principles density functional theory (DFT)
calculations within the local density approximation (LDA)\cite{lda} using
projector augmented wave (PAW) potentials.\cite{paw}. All results discussed
in the text are obtained using LDA. To compare with the LDA results of specific
systems, we also performed calculations using Generalized Gradient
Approximation\cite{pbe}(GGA) together with van der Waals (vdW) correction,
(GGA+vdW).\cite{grimme06} In our earlier tests LDA yielded
interlayer spacings of layered materials and other structural parameters in
agreement with experimental data, as well as with those obtained by using
GGA+vdW. In the vdW corrections of the later method, DFT description is restricted
to shorter correlation length scales, but for the medium and large inter-atomic
distances the damped $C_{6}r^{-6}$ term is used. The systems, whose numerical
values are also obtained by using GGA+vdW are indicated. Numerical calculations
are performed using VASP.\cite{vasp} Kinetic energy cutoff, $E_{cut}= \hbar^2
|\mathbf{k}+\mathbf{G}|^2 / 2m$, for plane-wave basis set is taken as 500 eV. A
vacuum spacing of at least 15 \AA~ is placed between adjacent layers to hinder
the layer-layer interactions. The convergence criterion of self consistent
calculations for ionic relaxations is $10^{-5}$ eV between two consecutive
steps. Structural optimizations were performed using a conjugate gradient
algorithm with a convergence criterion of $10^{-4}$ eV/\AA. Pressures on the
lattice unit cell are decreased to values less than 1.0 kBar. The adsorption of
a single Cl atom to graphene surface is treated using the supercell geometry,
where single Cl is adsorbed to each (4x4) supercell. In the self-consistent
potential and total energy calculations using a (4x4) supercell of
chlorographene, a set of (13x13x1) \textbf{k}-point sampling is used for
Brillouin zone (BZ) integration. The sampling of BZ is then scaled according to
the sizes of the supercells used for other systems. Ground state
electronic structures are calculated by
applying a dipole correction to eliminate the artificial electrostatic field
between periodic supercells. For the charge transfer analysis, the effective
charge on atoms are obtained by Bader method.\cite{bader}

The stabilities of structures having various Cl coverage are examined by the
calculation of phonon frequencies for \textbf{q}-wave vectors over BZ by using
both small displacement method (SDM)\cite{alfe} and density functional
perturbation theory (DFPT).\cite{pwscf} We used 196-atom supercell of
chlorographene, \textbf{q}-point sampling grid of 3x3x1 and 0.01 \AA~
displacements in calculations using SDM. DFPT part of the phonon calculations
were performed by using 6x6x1 grid of \textbf{q}-points for chlorographene
unitcell. A given structure is considered to be stable, if vibration frequencies 
are positive for all \textbf{q}-points in BZ.

The energy band gap, which is usually underestimated by DFT, is corrected by
frequency-dependent GW$_{0}$ calculations.\cite{gw} In GW$_{0}$  corrections
screened Coulomb potential W, is kept fixed to initial DFT value W$_{0}$ and
Green's function G, is iterated four times. Finally, the band gap of CCl is
calculated by using (12x12x1) \textbf{k}-points in BZ, $20$~\AA~ vacuum spacing,
default cut-off potential for GW$_{0}$, 160 bands and 64 grid points.

\begin{figure}
\includegraphics[width=8.5cm]{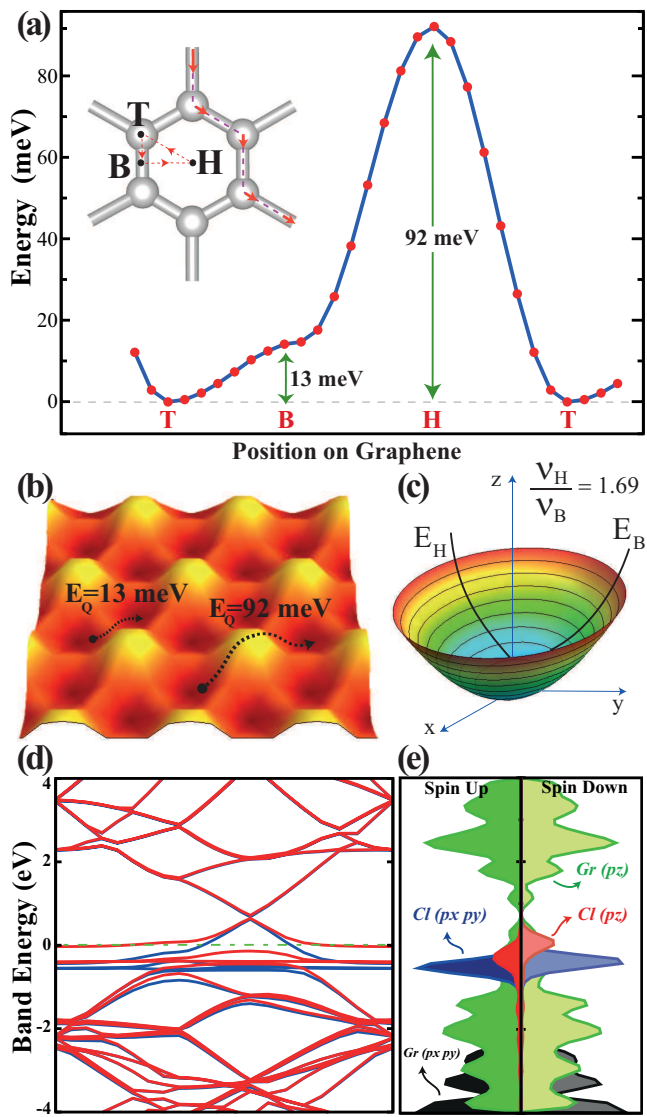}
\caption{\label{f1}(Color online) (a) Variation of energy for a single Cl adatom 
along the symmetry direction of a hexagon. Zero of energy is set to the energy of 
T-site. The diffusion path with the lowest energy barrier of $Q$=13 meV between 
two adjacent T-sites are marked with thick red/ dashed lines. (b) Energy landscape 
of a single Cl adatom adsorbed to graphene. Dark (light) colors represents the 
top (hollow) sites. (c) Potential energy contour plots (paraboloid) of Cl adatom 
around the T-site. The jump frequency of Cl atom $\nu$, for different directions 
are calculated from this paraboloid. (d) Band structure of a single Cl adsorbed 
to each (4x4) supercell of graphene and corresponding total (TDOS) and orbital 
decomposed (PDOS) densities of states. The zero of band energy is set to the 
Fermi level.}
\end{figure}

The binding energy of a single adatom X to the graphene supercell (i.e. X=H
hydrogenation; X=F fluorination; X=Cl chlorination) is calculated according to
the expression, $E_{b}=E_{T}[Gr]+E_{T}[X]-E_{T}[Gr+X]$, in terms of the ground
state total energies\cite{groundstate} of bare graphene $E_{T}[Gr]$, free X atom
$E_{T}[X]$, and X adsorbed to graphene $E_{T}[Gr+X]$. Accordingly, $E_b > 0$
indicates a bound state. Similarly, the formation energy of a single adatom 
adsorbed to graphene supercell relative to X$_2$ molecule is defined as
$E_{f}=E_{T}[X_{2}]/2+E_{T}[Gr]-E_{T}[Gr+X]$,
where $E_{T}[X_{2}]$ is the ground state total energy\cite{groundstate} of
X$_{2}$ molecule. For an exothermic process $E_{f} > 0$. The cohesive energy of
graphene fully covered with X, CX relative to free C and X atoms is defined as
$E_{coh} = 2E_T[X] + 2E_T[C] - E_T[CX]$, where $E_T[CX]$ is the total ground
state energy (per unit cell) of CX. By definition $E_{coh} >$ 0 indicates
binding relative to individual constituent atoms.

\section{Adsorption of Single Chlorine}\label{single}

Understanding of the adsorption process of a single Cl adatom on graphene is
essential for the investigation of its chlorinated derivatives. The Cl-Cl
coupling between adsorbed Cl atoms is crucial. Here, the Cl coverage is defined
as the ratio of the number of Cl atoms $N_{Cl}$ to the number of carbon atoms
$N_{C}$ in a supercell; namely $\Theta = N_{Cl}/N_{C}$. A single Cl atom
adsorbed to graphene is represented by a system where a Cl atom is adsorbed to
each (4x4) supercell. This actually corresponds to a uniform coverage of
$\Theta$=0.03125. Based on the analysis discussed in the next section we found
the size of (4x4) supercell is sufficient to neglect the Cl-Cl coupling. Even if
the Pauling scale of electronegativity of specific atoms alone cannot provide
criteria for the character and strength of the bonds between those atoms, it
usually indicates useful correlations. For example, according to the Pauling
scale of electronegativity (H: 2.20, C: 2.55, Cl: 3.16 and F: 3.98) the binding
of Cl atoms to graphene can be expected to be stronger than that of hydrogen
adatoms (0.98 eV) and weaker than the fluorine adatoms (2.71 eV). Three
different adsorption sites can be foreseen for the adsorption of Cl on graphene
as described in Fig.~\ref{f1}(a). These are the hollow (H) site above the center
of hexagon, the top (T) site on top of C atoms, the bridge (B) site above the
middle of C-C bonds. Among these sites, the strongest binding of single Cl to
graphene is attained at T-site with a binding energy of $E_b$=1.10 eV ($E_b$
calculated using GGA+vdW 1.16 eV). Earlier studies dealing with graphene-Cl
interaction calculated the binding energies of Cl adatom ranging 1.05
eV\cite{ijas} using ($4x4$) supercell and 0.8 eV.\cite{wehling} Using a
methodology similar to that used in the present work the binding energy of Cl
adatom is calculated to be 1.13 eV.\cite{jpc} Applying correction based on
hybrid functionals\cite{hse} and hence performing GGA+HSE06 calculations we
found the binding energy of Cl as 0.7 eV.

The variation of the total energy of an adsorbed Cl atom at the symmetry points
and along the symmetry (T-B-H-T) directions of a hexagon in the (4x4) supercell
is shown in Fig.~\ref{f1} (a). At each point on the energy curve, $x$- and
$y$-positions of adsorbed Cl atom are fixed, its $z$-height, as well as
positions of all C atoms in the (4x4) supercell are optimized by minimizing
total energy and atomic forces. Using this energy curve we reveal the energy
barriers to be overcame by the adsorbate migrating or diffusing on the surface
of graphene. The minimum energy barrier occurs at the B-site between two
adjacent T-site. The relevant energy barriers are shown on the calculated energy
landscaping presented in Fig.~\ref{f1} (b). This analysis suggests that a
diffusing Cl atom can take a path of minimum energy barrier following the edges
of hexagon from T-site to T-site through the barrier of $Q$=13 meV at B-site.
This energy barrier is in fair agreement calculated by Wehling et
al.\cite{wehling} We note that this barrier is very low and allows Cl atom to
migrate on the surface of graphene as if it rolls along the edges of hexagon.
This is a remarkable situation, where the migration or diffusion of Cl occurs
with almost no barrier, but it remains bound to the surface. This property of Cl
adatom on graphene heralds a number of possible applications, such as superlow
sliding friction, sensors and devices for energy harvesting.\cite{harvest} The
binding energy of an isolated Cl$_{2}$ molecule is calculated to be 3.58
eV,\cite{Cl2} leading to negative formation energy $E_{f}<$0. Then the
adsorption of chlorine on graphene is an endothermic process and hence it can be
achieved by, for example, the assistance of light absorption.\cite{bli}

\begin{figure}
\includegraphics[width=8.5cm]{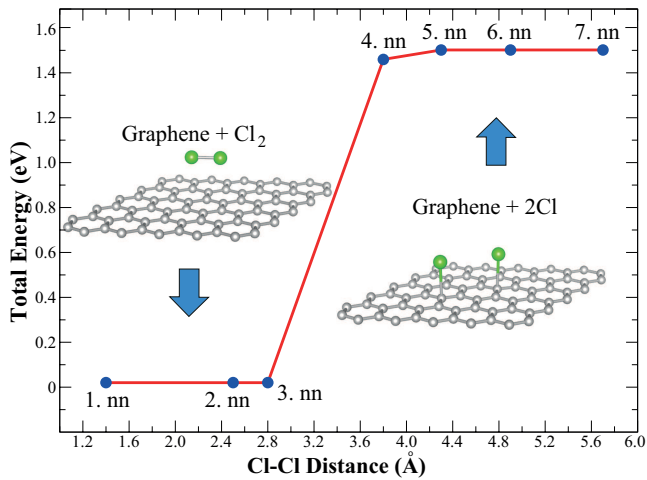}
\caption{(Color online) The interaction energy between two Cl atoms adsorbed to
the same side of a (6x6) supercell of graphene. The zero of energy is set to the
energy of Cl$_2$ plus graphene. nn denotes the nearest neighbor in graphene
lattice }
\label{f2}
\end{figure}

In Fig.~\ref{f1} (c) the energy plot $E_{T}(x,y)$ (or energy paraboloid) of Cl
adatom displacing from its equilibrium position at T-site is deduced
approximately from the energy landscaping shown in Fig.~\ref{f1} (b). When
displaced along T-H and T-B directions, the Cl adatom undergoes different energy
variations. Using this energy plot and the model related with simple harmonic
oscillator we estimate the values of jump frequencies of adatom towards H- and
B-sites to be $\nu_{T \rightarrow B} \sim$ 0.97 THz and $\nu_{T \rightarrow H}
\sim$ 1.68 THz, respectively. The diffusion constant at a given temperature can
be obtained from a simple expression,\cite{kittel} $D=\nu a^{2} exp(-Q/k_{B}T)$ 
in terms of energy barrier $Q$, lattice constant $a$ and the characteristic jump
frequency $\nu$. Using these values the diffusion constant at T=300 K is
calculated for the barrier $Q$=13 meV to be $D$=3.57 $cm^{2}/s$. This indicates
that Cl adatom migrates readily over the graphene layer.

The length of Cl-C bond is found to be $d_{Cl-C}$=2.54 \AA. This value is larger
than the bond length calculated from the sum of covalent radii of Cl and C
atoms,\cite{kittel} namely $r_{Cl}+r_{C}=1.75$\AA, as well as the bond lengths
of CCl$_4$ (1.77 \AA). This situation indicates that the character of this bond
is rather different from C-Cl covalent bond in  CCl$_4$. When single Cl is
adsorbed to graphene, 0.44 $e$ is transferred from graphene to Cl adatom
according to Bader analysis.\cite{bader} The calculated charge transfer, as well
as the analysis of charge density distribution indicates an ionic character for
the Cl-C bond.\cite{jpc} The negative charge on Cl induces a dipole moment of
\textbf{p}=0.67 $e\cdot$\AA~when a single Cl atom is adsorbed to each (4x4)
supercell of graphene.\cite{cohen-adsorp} Consequently, the work function of
bare graphene which is 4.49 eV increases to 6.53 eV upon Cl adsorption. This the
energy required to remove an electron from the Fermi level to the vacuum energy
at the side where Cl atoms are adsorbed. Incidentally, while the atomic
configuration of graphene beneath Cl adatom is rather flat, H and F adatoms on
graphene impose buckling resulting in a covalent
character.\cite{apl2009,hasancf} As we will discuss in the next section, charge
transfer from C to Cl and hence the character of the bonding in two sided
adsorption is dramatically different.

While single Cl adatom on the surface of graphene can be viewed as an impurity
leading to localized (or resonance) states, our model representing the
adsorption using a (4x4) supercell gives rise to an energy band structure, where
dispersion of bands related with adsorbate can be taken as the measure of
adsorbate-adsorbate coupling. The band structure of a single Cl adsorbed to each
(4x4) supercell of graphene presented in Fig.~\ref{f1} (d) has filled chlorine
states appearing as flat bands. In view of the negligible dispersions of these
bands associated with Cl and the analysis of the bands corresponding to (5x5)
and (6x6) supercells in the following section we concluded that a Cl adsorbed to
each (4x4) supercell can mimic successfully the single, \textit{i.e.} isolated
Cl adsorbed to graphene surface. Because of 7 valence electrons and hence a
single unfilled $3p$ orbital occurring below the Fermi level $E_F$ of bare
graphene, Cl adsorbed graphene is metalized and becomes magnetic with
$\mu$=0.56 $\mu_{B}$ due to broken spin degeneracy. The linearly crossing bands
of bare graphene is raised above $E_F$. The orbital decomposed density of states
in Fig.~\ref{f1} (e) show that the flat bands below $E_F$ are derived  mainly
from Cl-$3p_x$ and Cl-$3p_y$ orbitals.

\section{Coverage of Graphene by Chlorine Adatoms}\label{coverage}

The interaction between two adsorbed Cl as a function of their separation is
important to understand the coverage dependent properties and stability of Cl
covered graphene. To this end we consider two Cl atoms adsorbed to the (6x6)
supercell of graphene and calculate the total energies as a function of the
separation between them. Chlorine atoms are placed at specific adsorption sites
by fixing their $x$- and $y$-coordinates, but their $z$-coordinates, as well as
the positions of other C atoms in the supercell are relaxed. In Fig.~\ref{f2} we
show how the Cl-Cl interaction changes with their separation. The Cl-Cl coupling
is practically negligible when their separation greater or equal to 3.6 \AA~
corresponding to fourth nearest neighbor. For smaller separations corresponding
to third nearest neighbor separation of graphene with the threshold distance,
they form Cl$_2$ molecule
and desorb from graphene surface, since Cl-Cl interaction energy becomes
stronger than the sum of the binding energies of two Cl atoms to graphene (2.20
eV). Accordingly, the gain of energy through the formation of Cl$_2$ is 1.26 eV
per molecule. This explains also why chlorination of graphene is endothermic.
There is a weak vdW interaction between bare graphene and Cl$_2$ molecule. The
maximum interaction of 144 meV occurs when the molecule is parallel to a C-C
bond and 3 \AA~above it.

\begin{figure}
\includegraphics[width=8.5cm]{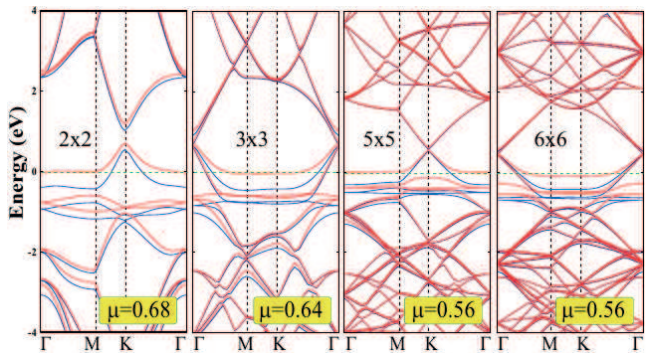}
\caption{(Color online) Energy band structure of a single Cl atom adsorbed to
each ($n$x$n$) supercell of graphene for $n$=2,3,5 and 6, which correspond to
the one-sided uniform coverage $\Theta=1/2n^{2}$. For $n \ge 2$ the distance
between nearest Cl adatoms is larger than the the threshold distance described
in Fig.~\ref{f2}. Whereas stable Cl coverage with $n=1$ (or $\Theta$=0.5) cannot
be achieved due to the strong Cl-Cl coupling. The units of magnetic moments
$\mu$ is Bohr magneton per ($n$x$n$) supercell. The zero of energy is set at the
Fermi level, $E_F$.}
\label{f3}
\end{figure}

Next we examine how the electronic structure changes with one sided uniform
coverage. In Fig.~\ref{f3} we show the band structures calculated for single Cl
atom adsorbed to ($n$x$n$) supercell for $n$=2,3,5 and 6. Here $n$=2 and $n$=3
correspond to $\Theta$=0.125 and 0.056. The phonon calculations for these two
coverage values are found stable, since the frequencies of phonon modes are
positive for any \textbf{k}-point in BZ. The smallest separation between Cl
atoms is larger than the threshold distance of 2.8~\AA~(i.e. the third
nearest neighbor distance) even for $\Theta$=0.125.
We see that with increasing $\Theta$ the dispersion of Cl bands increases
slightly and the linearly crossing bands start to split. The magnetic moment
also increases with coverage, since the splitting of spin-up and spin-down bands
increases. We found $\mu$=0.64 $\mu_B$ and  $\mu$=0.56 $\mu_B$ for $\Theta$=0.0555 
(or $n$=3) and $\Theta$=0.020  (or $n$=5), respectively. On the other hand, a single Cl
adsorption to each (1x1) unit cell of graphene (which corresponds to
$\Theta$=0.5) is unstable due to Cl-Cl distance smaller than 2.8~\AA. Since the
uniformly chlorinated graphene is found to be metallic for any coverage studied
here, the experimental studies measuring a band gap of $\sim$0.05 eV should
correspond to two sided coverage.

\begin{figure}
\includegraphics[width=8.5cm]{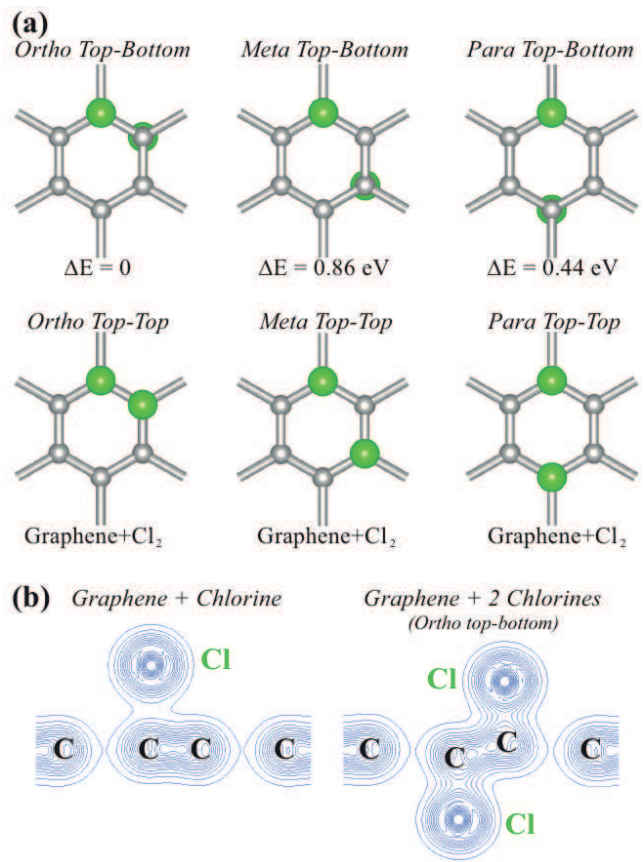}
\caption{(Color online) (a) The atomic structure two Cl atoms adsorbed to a
(4x4) supercell of
graphene. In three different configuration illustrated by top panels, namely
ortho top-bottom, para top-bottom and meta top-bottom, two adsorbed Cl atoms are
stable. $\Delta E$ indicates their energies relative to the total energy of the
ortho top-bottom configuration. Double sided adsorption imposes a local buckling
in planar graphene. Three one-sided configurations, ortho top-top, para top- top
and meta top-top are not allowed; Cl atoms cannot be bound to graphene, they
rather form Cl$_2$ molecule. Large green and small gay balls represent Cl and C
atoms, respectively. (b) Contour plots of the total charge density of a single
Cl-C bond and two Cl-C bonds in ortho top-bottom configuration. Contours
spacings between 0.025$e$/\AA$^{3}$ and 1.0 $e$/\AA$^{3}$ are 0.025
$e$/\AA$^{3}$.}
\label{f4}
\end{figure}

To understand the formation of chlorinated domains on graphene, and hence to
analyze the kinetics of Cl coverage, we next examine the adsorption of second Cl
atom at different sites at the close proximity of the first one in (4x4)
supercell. When a Cl is placed on top of a carbon atom, second Cl can be
adsorbed on six relevant sites on the hexagonal carbon ring as shown in
Fig.~\ref{f4} (a).

Among the various possible cases, the ortho top-bottom arrangement is the most
favorable one. While adsorption of a single Cl atom yields 0.05 \AA~ buckling of
graphene lattice, adjacent C atoms forming C-Cl bonds in ortho top-bottom
configuration are buckled by 0.46 \AA. In Fig. \ref{f4}(b) the contour plots of
total charge density on the planes passing through the bonds clearly explain the
ionic character of C-Cl bond of a single Cl adatom. In the case of ortho
top-bottom the bonding between Cl adatom and C atom of graphene has
$sp^{3}$-type covalent character. While the charge density between Cl and C
atoms is very low in the ionic case, the bond charge is enhanced in the covalent
C-Cl bond. The binding energy of covalent bond is stronger ($E_b$=1.53 eV) and
consequently its length is shorter ($d$=1.90 \AA, which is comparable with the 
empirical covalent radii of the C-Cl bond or the bond length $d_{C-Cl}$=1.77 \AA~
of CCl$_4$ molecule). The binding energies per bond
in the case of meta and para top-bottom configurations are $E_b$=1.10 eV and
$E_b$=1.31 eV, respectively. In spite of the fact that the ortho top-bottom
configuration is 0.52 eV less energetic than the formation of Cl$_2$ molecule,
it can be stable since the formation of Cl$_2$ molecule from two Cl atoms at
different sides are hindered. In para and meta top-bottom structures, the
binding energies of two bonds are further lowered from the formation energy of
Cl$_2$ molecule.

The situation is however dramatically different for one sided (top-top)
configurations 
presented in the second row of Fig. \ref{f4} (a). For these three one-sided
configuration, namely ortho, para and meta top-top configurations the Cl-Cl
distances are within the third nearest neighbors. Owing to the strong Cl-Cl
coupling, the formation of  Cl$_2$ molecule is energetically favored.
Accordingly, unlike the half fluorinated graphene\cite{hasancf} (C$_2$F),
one-sided densely chlorinated graphene with the coverage $\Theta$=0.5, namely
C$_2$Cl, cannot be realized.

\begin{figure}
\includegraphics[width=8.5cm]{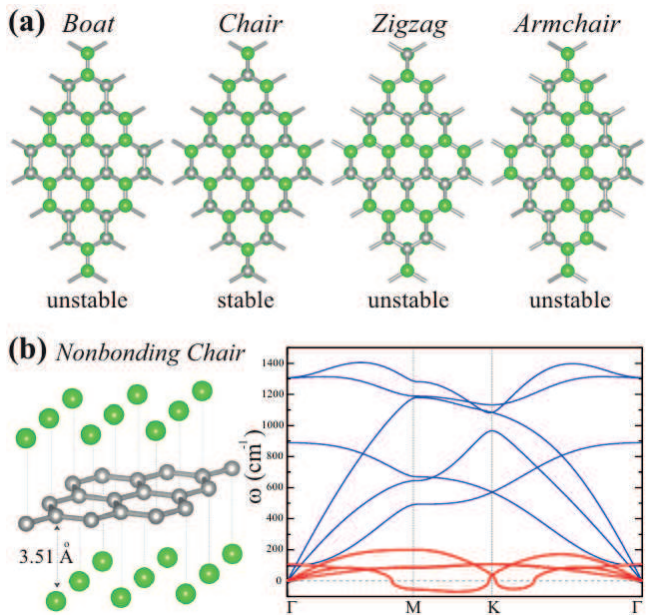}
\caption{(Color online) (a) Top view of atomic structures of boat, chair (\textit{i.e.}
covalently bonded and buckled graphene layer is sandwiched between two planar Cl layers),
zigzag and armchair conformations. Green and small gay balls represent Cl and C atoms,
respectively. (b) Side view of nonbonding chair conformation consisting of
one planar graphene layer sandwiched by two planar Cl layers and its calculated
phonon dispersion curves. Low frequency phonon modes shown by red lines are
related to adsorbed Cl atoms. These modes have imaginary frequencies and hence
they are unstable.}
\label{f5}
\end{figure}

We now concentrate on the chemical conversion of graphene to a fully chlorinated
graphene structure ($\Theta$=1.0), called chlorographene. In addition to
well-known boat, chair and nonbonding chair conformations one can also consider
zigzag and armchair stoichiometric chlorographene configurations shown in Fig.
\ref{f5}. Here, boat, chair, nonbonding chair and zigzag conformations are
treated using (2x2) supercell; a (4x4) supercell is required for the armchair
conformations. As a consequence of the strong Cl-Cl interaction, boat, zigzag
and armchair configurations of Cl atoms cannot remain stable on graphene.

Here the nonbonding chair conformations in Fig. \ref{f5} (b) consists of a
planar graphene layer sandwiched between two Cl layers deserves a detailed
discussion. In this conformation Cl atoms are placed to the alternating T-sites
at both sides of graphene. Earlier theoretical studies\cite{klintenberg,
medeiros} have predicted that this metallic structure has highest cohesive
energy among conformations described in Fig. \ref{f5}. Present calculations
predict $E_{T}$=23.35 eV/per primitive cell and cohesive energy $E_{coh}$=2.79
eV/ per primitive cell. Even if the structure optimization using conjugate
gradient method performed in the (1x1) hexagonal unit cell finds the nonbonding
chair structure stable, we carried out an extensive analysis of this
conformation. We found that the attractive interaction between each Cl layer and
underlying graphene is rather small (50 meV per unit cell) in spite of the fact
that graphene lattice expanded by 2.8\% and the system attained a total energy
$\sim$3.31 eV higher than that of graphene. The weak interaction between Cl and
C atoms explains why the distance from Cl layer to graphene is large (3.51 \AA).
This large distance and minute charge transfer from carbon to Cl atom can be
contrasted with the C-Cl distance (2.54 \AA) and the transfer of 0.44 electrons
from C to Cl of the ionic bond of single Cl atom adsorbed to graphene as
discussed in Sec. III. This paradoxical situation occurred due to the strong
intra-layer coupling among Cl atoms in both Cl-layers at both sides of
graphene. In fact, the binding energy of Cl atoms of a single Cl layer having
the same atomic configuration as the Cl layers of nonbonding chair conformation
is found to be 1.41 eV per Cl atom. It appears that high cohesive energy of
nonbonding chair conformation relative to free graphene and free Cl atoms is
attained mainly by the cohesion between Cl atoms, but not by the interaction
between Cl and carbon atoms. In addition, GGA+vdW calculations performed for
nonbonding conformation resulted with optimized structure close to that obtained
by using LDA.

Under these paradoxical situations we further examined the stability of the
metallic nonbonding chair conformer by carrying out ab-initio calculations of
phonon frequencies in the BZ. As seen in Fig. \ref{f5} (b), the calculated
phonon branches, which are associated with Cl in-plane and out-of-plane modes
and hence are practically isolated from graphene modes, have imaginary
frequencies near $M$- and $K$-high symmetry points. This clearly indicates that
nonbonding chair structure is unstable at T=0 K. In fact, a minute displacements
or perturbation of atoms in Cl plane is resulted in the breakdown of Cl-layers,
if nonbonding chair structure is treated using a (4x4) supercell. It appears
that structure optimization in earlier studies\cite{klintenberg, medeiros}
carried out in a single cell has limited the degree of freedom of Cl atoms
preventing them from reconstructions involving large displacements or from
forming molecules. Accordingly, the stability imposed by the structure
optimization using primitive cell was unrealistic. Even though the nonbonding
chair conformation has the lowest energy among other conformers including
(covalently bonded) chair structure in Fig. \ref{f5}, it corresponds either to a
very shallow local minimum or to a saddle point for specific directions of
atomic displacements in Born-Oppenheimer surface.

\begin{table*}
\caption{Calculated values for graphene, and graphene derivatives, such as
graphane CH, fluorographene CF and chlorographene CCl. These are lattice
constant ($a$); C-C bond distance ($d_{CC}$); C-X bond distance ($d_{CX}$);
thickness of the layer ($t$); photoelectric threshold ($\Phi$); charge transfer
from C to X ($\Delta \rho$); cohesive energy per unit cell ($E_{coh}$),
formation energy ($E_f$), desorption energy of a single X atom from the CX
surface ($E_{des}$); direct band gap ($E_{g}$); band gap corrected with GW$_o$,
$E_{g}^{GW_{o}}$ ; in-plane stiffness ($C$); Raman active modes. Energies are
calculated in (2x2) supercell.}
\label{table}
\begin{center}
\begin{tabular}{cccccccccccccccccc}
\hline  \hline
&$a$ & $d_{CC}$& $d_{CX}$ & $t$ & $\Phi$ & $\Delta \rho$ &  $E_{coh}$ &$E_{f}$ & $E_{des}^{X}$ & $E_{g}$ & $E_{G}^{GW_{o}}$ &$C$& \textbf{R}-$active$ $Modes$\\

Material (CX) & \AA&\AA  &\AA  & \AA & $eV$ & $e$ & $eV$ & $eV$ & $eV$ &$eV$  &$eV$ &$J/m^{2}$&$cm^{-1}$\\
\hline
Graphene & 2.46 & 1.42 & -& - & 4.49 & - &  17.87 & - & -  & -&- &  335& 1600\\
 \hline
CH & 2.51 & 1.52& 1.12& 2.68 & 4.79 &  0.06 & 23.60 & +0.39&  4.79 &3.42 &5.97 &  243&   1162, 1164, 1341, 2806\\
 \hline
CF & 2.55 & 1.55& 1.37& 3.22 & 7.93 & -0.61 & 25.31 & +2.04&  5.46 &2.96 &7.49 & 250& 245, 681, 1264, 1305\\
 \hline
CCl& 2.84 & 1.72&1.73 &3.96 & 3.67 & -0.13 & 19.60 & -0.95&  1.28 & 1.21 &4.33 & 186&105, 398, 715, 1042 \\

\hline \hline
\end{tabular}
\end{center}
\end{table*}

\section{Stable Fully Chlorinated Graphene: Chlorographene} \label{full}

\subsection{Structural Properties}\label{structural}

\begin{figure}
\includegraphics[width=8.5cm]{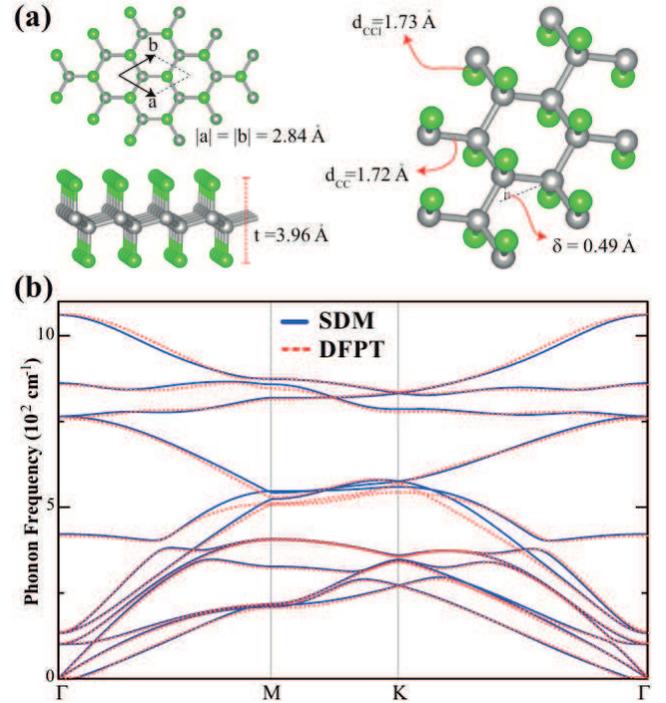}
\caption{(Color online) (a) Top, side and tilted views for the atomic structure
of chlorographene CCl layer having hexagonal lattice and honeycomb structure.
Carbon and chlorine atoms are indicated by gray (dark) and green (light) balls,
respectively. Calculated structural parameters are indicated. (b) Phonon bands
of chlorographene calculated using by SDM\cite{alfe} and DFPT\cite{pwscf}
methods.}
\label{f6}
\end{figure}

In contrast to nonbonding chair conformation, our analysis indicates that like
graphane (CH) and fluorographene (CF), the chair structure, where one chlorine
atom is
attached to each carbon atom of buckled graphene alternatingly from top and
bottom sides can be a stable structure. The optimized atomic structure and the
hexagonal lattice are shown in Fig.~\ref{f6} (a). The present study predicts its
total energy $E_T$=22.24 eV/per primitive cell and its cohesive energy
$E_{coh}$=1.68 eV. In the rest of paper we name this stable structure as
chlorographene or CCl, which should be distinguished from the unstable,
nonbonding chair structure presented in the previous section in Fig.~\ref{f5}
(b). The remarkable situation is that the $sp^2$-bonded planar graphene is
buckled by $\delta$=0.50 \AA, once it is converted to CCl. This way the local
configuration of three folded $sp^2$-bonding has changed to four folded
$sp^3$-bonding, reminiscent of tetrahedrally coordinated diamond structures.
Formation of four folded $sp^3$-bonding through buckling is essential for the
stabilization of CCl, despite the C-C distance increased from 1.42 \AA~to 1.72
\AA. On the other hand, while the C-Cl ionic bond in the adsorption of single Cl
adatom is 2.54 \AA~in Sec.III, it transforms to a covalent bond by contracting
to 1.73~\AA ~in chlorographene, which is also very close to the bond length of
CCl$_4$ molecule. The bond length of chlorographene can also be contrasted with
the Cl-graphene distance in nonbonding chair structure discussed in the previous
section. We note that chlorographene is stable despite its cohesive energy is
smaller than that of the unstable nonbonding chair conformation. This is due
the fact that chlorographene corresponds to a local minimum in Born-Oppenheimer
surface.

In Table \ref{table} we compare the structural parameters, relevant energies
like cohesive energy, formation energy and desorption energy of graphene, CH, CF
and CCl, all calculated by using LDA. It is seen that chlorographene, CCl has
binding energy $E_{b}$=1.68 eV per unit cell relative to bare graphene and free
Cl atom, and lowest $E_{coh}$ (19.60 eV) per unit cell among other possible
graphene derivatives. However, once CCl is synthesized a desorption energy of
$E_{des}=$1.28 eV is required to take a single Cl atom out.

\subsection{Vibrational Properties and Raman Spectra}\label{vibrational}

Chlorographene (CCl) posses D$_{3d}$ point group symmetry. Phonon bands of CCl
is shown in Fig.~\ref{f6}(b). It is seen that all the phonon modes have positive
frequencies and hence the predicted structure of chlorographene is stable at T=0
K. Unlike the phonon bands of nonbonding conformer, C-Cl bonds of chlorographene
participate to all acoustic phonon branches. Therefore, the phonon spectrum of
chlorographene differs from the spectrum of graphene. Though the chlorographene
belongs to the same space group with CH and CF,\cite{hasancf, leenaerts} the
phonon frequencies are lowered (softened) due to the saturation of C atoms with
heavy Cl atoms. While LA and TA modes have linearly dependent to phonon wave
vector, ZA mode has a quadratic dispersion in the vicinity of
$\Gamma$-point.\cite{jia, hasancf, leenaerts} As compared to single layer
graphene, LO and TO (ZO) optical modes are softened from 1600 (900) to 1061
(421) $cm^{-1}$ due the existence of surrounding Cl layers. In general, both
phonon bands calculated using SDM and DFPT agree well, but they slightly differ
at M and K point for optical phonon branches near at 500 $cm^{-1}$.

Group theory analysis shows that the decomposition of the vibration
representation at the $\Gamma$-point is $\Gamma = 2A_{1g} + 2A_{2u} + 4E_{g} +
4E_{u} $. Among these, the modes at 105, 398, 715 and 1042 $cm^{-1}$ are bond
stretching modes and are Raman-active. Raman mode $A_{1g}$ at 1042 (398) is
entirely due to the out-of-plane vibration of C and Cl atoms moving in the same
(opposite) direction with respect to each other. The observation of these Raman
active modes are expected to shed light on the Cl coverage and the structure of
chlorinated graphene. The observation of characteristic D-peak at 1330, G-peak
at 1587 and 2D-peak at 2654 $cm^{-1}$ from chlorinated graphene indicates low
coverage of Cl.\cite{bli} Since the Raman measurements were performed in the
range of 1250-3500 $cm^{-1}$, possible Raman-active peaks
originating from chlorine atoms could not be observed.

\begin{figure}
\includegraphics[width=8.5cm]{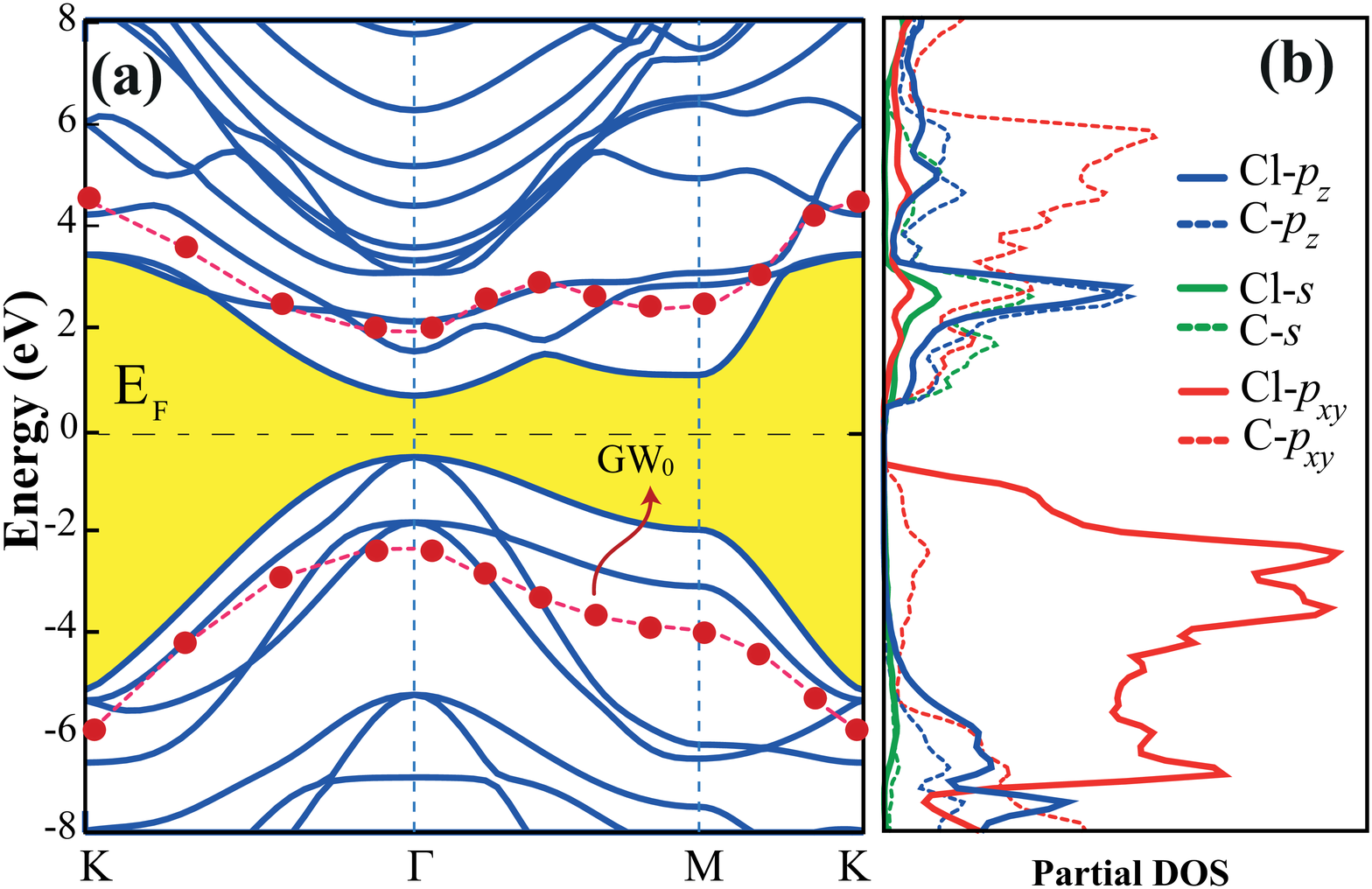}
\caption{(Color online) (a) Electronic band structure of chlorographene CCl.
The band gap is shaded yellow. The GW$_o$ corrected valance and conduction bands
are shown by dashed line and red balls. The zero of energy set to the Fermi
level $E_F$. (b) Density of states projected to various orbitals (PDOS).}
\label{f7}
\end{figure}

\subsection{Stability of CCl at finite temperature}

Here we investigate whether CCl is stable at finite temperature, even though all
phonon modes having positive frequencies in the Brillouin zone  indicate its
stability at T=0 K. This is achieved by carrying out ab-initio molecular
dynamics (MD) calculations at finite temperatures using $(6x6)$ supercell
involving 144 C and one Cl atoms. Here we summarize our findings. At relatively
low temperatures, for example at T=500 K, the perfect CCl remained stable even
after $\sim$6000 time steps of 2 fs. Even if 6000 time steps large for ab-initio
calculations, but low to attain a reliable statistics, this result suggests the
stability of perfect CCl near room temperature. However, Cl atoms dissociate
from CCl at 1000 K. This may be interpreted that CCl cannot be stable at
elevated temperatures. Ab-initio MD calculations with single Cl vacancy at one
side brought about the formation of a second vacancy at the other side after 200
time steps at 500 K. Thereafter, the structure continued to be stable. However,
an island of Cl at both sides of graphene, for example, 6 Cl atoms adsorbed to a
hexagon of graphene alternatingly from the top and bottom side of graphene is
found unstable.

Our analysis using finite temperature ab-initio MD calculations led us to draw
the following conclusion from the above results: A perfect CCl is stable at T=0
K and can remain stable possibly at room temperature. The creation of a single
vacancy at one site imposes the formation of a second vacancy at the opposite
side. This pair of vacancy can survive at room temperature. However, CCl having
vacancies or holes is vulnerable to dissociation through the formation of Cl$_2$
molecules. The negative formation energy underlies these instabilities.

\subsection{Electronic Properties}\label{electronic}

We present the band structure and the orbital decomposed density of states of
fully chlorinated graphene CCl in Fig.~\ref{f7}. As a consequence of transition
from $sp^2$- to of $sp^{3}$-type bonding and the mixing with Cl orbitals through
Cl-C covalent bonds, linearly crossing semimetallic bands of graphene changes
dramatically and turn to a nonmagnetic semiconductor with a direct band gap of
$E_{g}=$1.21 eV (GGA+vdW value: 1.55 eV) at the center of BZ. The two-fold
degenerate bands at the top of valence band are mainly composed of the $p_{xy}$
valence orbitals of Cl and C atoms and belong to the E$_{g}$
irreducible representation. However, non-degenerate conduction band edge is
formed by the hybridization of C-$p_{xy}$, C-$p_{z}$, C-$s$ and Cl-$p_{z}$ and
belongs to the A$_{2u}$ irreducible representation. Calculated values relevant
for the electronic properties of CCl are given together with those of
graphene, CH and CF in Table \ref{table}. CCl has the lowest band gap (1.21 eV)
among graphene derivatives, i.e. graphane ($E_{g}=$3.42 eV) and fluorographene
($E_{g}=$2.96 eV). Our calculations also reveals that the experimentally
observed energy gap of 0.045 eV may arise from the partially chlorinated
regions. Since DFT usually underestimates the band gaps, the band gap of 3D bulk
crystals are successfully corrected by GW$_{o}$ self-energy method. In the
present paper we apply GW$_{o}$ method to correct valance and conduction bands
and hence the band gap of CCl. We found the corrected band gap 4.33 eV. This is
a dramatic increase. Similar situation occurred for the correction of band gaps
carried out for single layer structures, such as BN\cite{mehmet},
CF\cite{hasancf} and MoS$_2$\cite{can}. In particular, while LDA predicted band
gap of single layer MoS$_2$ agrees with experiment, GW$_o$ correction yielded
very large band gap. This situation led us to question whether GW$_o$ correction
is suitable for single layer structures.

While graphane have a positively charged surface due to the electron transferred
from H to C atoms, Cl layers of CCl are negatively charged since 0.13 electrons
(GGA+vdW value: 0.10 electrons) are transferred from C to Cl atoms. We also note
that the effective charge is reduced from 0.42 to 0.13 electrons by going from
ionic C-Cl bond in single Cl adsorption to the covalent C-Cl  bond of
chlorographene. Since CCl surfaces are negatively charged, it is possible to
lower the photoelectric threshold of graphene from 4.49 eV to 3.67 eV by
covering its surfaces with chlorine atoms. In contrast, as one notices in Table
\ref{table}, both fluorinated\cite{hasancf} and hydrogenated\cite{apl2009,
leenaerts} derivatives of graphene have photoelectric thresholds higher than
graphene and CCl. The lower photoelectric threshold provide materials highly
emissive surfaces. This facile photoemission feature is desirable for fast laser
applications.\cite{allen,laserop}

\subsection{Mechanical Properties}\label{mechanical}

Earlier studies have shown that graphane\cite{graphane-m,prb2010} and
fluorographene\cite{hasancf,leenaerts} derivatives are strong materials like
graphene. Among the three graphene derivatives chlorographene is the thickest
one and we can expect some different mechanical properties. We will discuss the
structural rigidity of chlorographene within the harmonic range of the elastic
deformation, where the structure responded to strain $\epsilon$ linearly and
reversibly. In the elastic range, in-plane stiffness $C$ would be a good measure
of the response of material. Here we use the expression to calculate in-plane
stiffness,  $C=(1/A_{0})\cdot(d^{2}E_{S}/d\epsilon^{2})$, where $A_{0}$ is the
equilibrium area of the supercell and the strain energy is defined as the total
energy at a given uniform strain minus the total energy at zero strain, namely
$E_{s}=E_{T}(\epsilon)-E_{T}(\epsilon=0)$. Here the strain in one direction is
$\epsilon=\Delta c/c_o$, where $c_{0}$ is the equilibrium lattice constant of
supercell and $\Delta c$ is its stretching. For CCl we calculated $C$ as 186
$J/m^{2}$ which identifies it as strong as CH and CF.

\begin{figure}
\includegraphics[width=8.5cm]{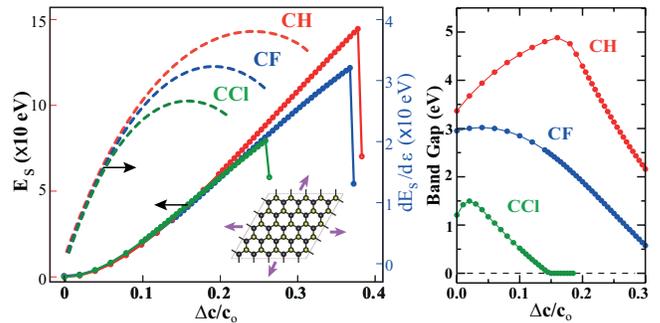}
\caption{(Color online) (a) The variation of the strain energy $E_s$ (curves at
the right hand side) and its derivative with respect to applied uniform strain
$\epsilon$, i.e. $d E_{s}/d \epsilon$, (curves at the left hand side) are
calculated for CCl, CH and CF. After the maxima these structures become unstable
and undergoes a plastic deformation. (b) Variation of the band gap with uniform
strain. Calculations performed in (5x5) supercells.}
\label{f8}
\end{figure}

In Fig. \ref{f8}(a) we show the variation of the strain energy $E_s$ and its
derivative with respect to the applied uniform strain, $dE_{s}/d\epsilon$. The
latter is linear for small $\epsilon$ in the harmonic range. The elastic
deformation occurs until the maximum of $d E_{s}/d \epsilon$, whereby the
structure attains its initial state when the applied strain is lifted. Beyond
the maximum the structural instabilities sets in with irreversible deformations.
The region beyond the maximum is called plastic region. It seen that CCl
undergoes instability, under 0.15 expansion of the lattice. Whereas, the
critical expansion values are 0.20 and 0.24 for CF and CH, respectively.

We also calculated the effect of elastic strain on the band gap of CCl, CF and
CH and presented our results in  Fig. \ref{f8}(b). CH and CCl have similar
response to elastic strain: The increase of their band gaps for small strain is
followed by a rapid decrease for large strain. In contrast, initially the band
gap of CF does not show any significant increase with increasing strain; for
small strain it is almost unaltered, but decreases rapidly for large strain.
Because of its smaller band gap, the metalization of CCl occurs earlier in the
course of expansion.

\section{Conclusions}\label{conc}

Motivated with the recent work by Li et al.\cite{bli}, who achieved the
photochemical chlorination of a bare graphene, we performed a first principle
study on the chlorination of graphene starting from a single Cl adatom
adsorption to fully chlorinated graphene CCl. We found that even if a Cl atom
can be bound to graphene with a significant binding energy, it can migrate on
the surface of graphene almost without an energy barrier. Formation of a Cl$_2$
molecule from two individual, migrating adatoms is energetically favorable when
they are at close proximity. Therefore the formation energy of the adsorption of
a single Cl atom relative to Cl$_2$ molecule is negative. In this respect, all
possible configurations or decoration obtained by the chlorination of one
surface of graphene and considered in this study are not stable. On the other
hand, the configuration, where two Cl atoms are adsorbed from opposite sides of
graphene to two adjacent carbon atoms forming $sp^3$-type covalent C-Cl bonds
and hence inducing buckling are stable, since the dissociation of these two Cl
atoms by forming Cl$_2$ molecule is hindered. Once the bare graphene is fully
chlorinated from both side inducing buckling of carbon atoms, the resulting
conformation named as chlorographene, is stable at T=0 K and also possibly at
room temperature. With its 1.21 eV direct band gap, stiff mechanical properties
and response to homogeneous strain this material displays interesting properties
for future technological applications. On the other hand, the nonbonding chair
conformation has cohesive energy higher than that of chlorographene in the chair
conformation and is found to be unstable. While both structures are not in
ground state, chlorographene corresponds to a local minimum on the
Born-Oppenheimer surface. Of course, a state having the energy lower than those
of both conformation is Cl$_{2}$ molecules, which are weakly attached to bare
graphene.

\section{Acknowledgements}

The authors thank Dr M. Topsakal for his assistance regarding the stability of
chlorographene and Mr O. Ozcelik for his assistance in specific calculations.
This work is supported by TUBITAK through Grant No:108T234. Part of the
computational resources has been provided by TUBITAK ULAKBIM, High Performance
and Grid Computing Center (TR-Grid e-Infrastructure). S. C. acknowledges the
partial support of TUBA, Academy of Science of Turkey.


\begin{thebibliography}{99}
\bibitem{groundstate} The total energy of magnetic states or non-magnetic states, whichever is lower, are used in calculating binding, cohesive and formation energies.
\end{thebibliography}
\end{document}